# Reluctant Pioneer of Nuclear Astrophysics: Eddington and the Problem of Stellar Energy

Helge Kragh[*]

**Abstract:** During the years from 1917 to 1921, A.S. Eddington was intensely occupied with Einstein's general theory of relativity and the epic eclipse expedition which confirmed one of the theory's predictions. During the same period, he investigated the old problem of why the stars shine, which led him to suggest two different subatomic mechanisms as the source of stellar energy. One of them was the annihilation of matter and the other the building-up of helium from hydrogen. This paper is concerned with Eddington's work in this area, a line of work to which he returned on and off during the 1920s but then abandoned. His decision to stop working on the stellar energy problem coincided with the first attempts to understand the problem in terms of nuclear physics and quantum mechanics. Why did Eddington not follow up his earlier work and why did he ignore the contributions of the nuclear physicists which in the late 1930s resulted in the first successful theories of stellar energy production?

## 1. The problem of solar energy

During the late Victorian era most physicists and astronomers agreed that ultimately the source of the Sun's energy was gravitational. The celebrated Helmholtz-Thomson (or Helmholtz-Kelvin) theory was first proposed by Hermann Helmholtz in a lecture of 1854 and subsequently developed into mathematical details by William Thomson, who in 1892 became Baron Kelvin. According to this theory the Sun contracted slightly and as a result lost gravitational potential energy, which was transformed into light and heat. From about 1865 to 1905 the Helmholtz-Thomson contraction theory commanded great authority and was generally recognised to be true. Although there were several alternatives to it, these did not disturb the theory's status as the standard theory of solar energy production [Kragh, 2016; Tassoul and Tassoul, 2004, pp. 67-72].

[*] Niels Bohr Institute, University of Copenhagen. E-mail: helge.kragh@nbi.ku.dk. This is a revised version of an invited paper given at a conference in Paris 27-29 May 2019 on "Arthur S. Eddington: From Physics to Philosophy and Back Again."

Despite its popularity the contraction theory was plagued by several problems, the most serious one of which was the short lifetime of the Sun – and consequently also of the Earth. Thomson found that the Sun could have existed as a strongly luminous body for only 20 million years, which was far less than the time-scale needed by geologists and evolutionary biologists. In addition to the time-scale problem, the basic assumption of a shrinking Sun lacked confirmation. Thomson calculated that the Sun's radius would decrease with 35 m per year, but astronomers could find no trace of the predicted diminution of the size of the Sun. Generally, the theory was difficult to test if testable at all. At about 1900 a growing number of scientists lost confidence in the theory but in most cases without abandoning it. They could find no better theory to replace it.

One alternative was widely discussed, though, namely that the Sun's energy might be explained on the basis of radioactivity and subatomic energy. The large amount of helium in the Sun's atmosphere suggested that radium and other radioactive elements were abundant in the interior of the Sun and might provide an answer to the puzzle of solar energy [Kragh, 2016]. The radioactive hypothesis was popular and enjoyed support by leading physicists and astronomers, among them Ernest Rutherford, Johannes Stark, Frederic Soddy, and George Howard Darwin. However, it was not developed beyond a qualitative and rhetorical level, and it lacked observational support even more than the contraction hypothesis. After a decade of discussion, it was realised that radioactivity was not after all a viable alternative to gravitational contraction. When the American astronomer Charles Abbot reviewed the solar energy problem in 1911, he was ambivalent and unwilling to disregard radioactivity as a possible cause. But in the end he preferred the contraction hypothesis over the radioactive hypothesis, finding it more satisfactory "to account for the solar heat by known causes rather than to invoke radio-activity of undiscovered materials" [Abbot, 1911].

Eddington was not concerned with the question of solar energy before World War I, but in one of his earliest scientific papers he commented on the related and much-discussed question of the age of the Earth. "Now the discovery of radio-activity has changed our ideas altogether," he wrote. "The assumptions on which Lord Kelvin's calculations rested are known to be untrue. We can assign no upper limit to the age of the Earth, and the thousands of millions of years demanded by geologists are freely conceded" [Eddington, 1906; Burchfield, 1975]. Eddington later returned to the question, but now focusing on the ages of both the Earth and the Sun and armed with the new nuclear theory of the constitution of atoms. In an address of 1920, he commented: "If the contraction theory was proposed to-day as a novel hypothesis I do not think it would stand the smallest chance of acceptance. … Only the inertia of tradition keeps the contraction theory alive – or, rather, not alive, but an unburied corpse" [Eddington, 1920]. He further stated that, "Lord Kelvin's date of the creation of the sun is treated with no more respect than Archbishop Ussher's," a reference to James Ussher's notorious conclusion from 1650 that God had created the world on the 22d October 4004 BC.[1] However, still in the 1920s not all astronomers agreed that the Helmholtz-Thomson theory was a corpse, buried or unburied.

## 2. Annihilation and fusion hypotheses

Although the Bohr-Rutherford nuclear theory of atomic structure came to revolutionise physics as well as astronomy, at the time of the Great War it was of little concern to astronomers. What mattered was that the atom was somehow a conglomerate of positive and negative charges often described collectively as "electrons." This kind of picture was suggested as early as 1901 by James Jeans, who speculated that the positive and negative electrons might "rush together and annihilate one another" [Jeans, 1901]. Jeans also used the idea in a paper of 1904, and many years later, after the positive electron had become a proton, he returned to the annihilation hypothesis in the form

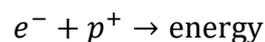

$$e^- + p^+ \to \text{energy}$$

According to Jeans, annihilation processes of this kind were important in astrophysics and cosmology [Jeans, 1924; Shahiv, 2009]. The same kind of process was considered by Eddington in 1917. He most likely knew about Jeans' old papers but encountered the hypothetical process in an even earlier work, namely in Joseph Larmor's *Aether and Matter* from 1900 [Eddington, 1926a].[2]

In his paper of June 1917, Eddington repeated his objections to the contraction theory, which he argued was irreconcilable with what was known about giant stars. As a way out of the short time-scale problem he appealed to "some unknown supply of energy," which might perhaps be "a slow process of annihilation of matter (through positive and negative electrons occasionally annulling one another)" [Eddington, 1917]. He most likely thought of the positive electrons as hydrogen ions or protons, but may have had in mind also annihilation of several electrons and nuclei heavier than hydrogen.

---

[1] Eddington apparently liked the Kelvin-Ussher comparison, which he reused in [Eddington, 1923] and also alluded to in [Eddington, 1927-28].

[2] Larmor formulated his idea of annihilation of positive and negative electrons within the framework of the theory of ether tubes and vortices. Eddington found the old ether picture appealing as an illustration and used it on several occasions [Eddington, 1926a; Eddington, 1926b, p. 294; Eddington, 1935, p. 166]. Also Jeans' early speculations on annihilation relied on the ether picture. See [Bromberg, 1976] for the earliest ideas of pair annihilation.

Eddington's cautious suggestion was quickly criticized by Jeans, who in spite of his early interest in the annihilation process maintained that some modified version of the contraction theory was preferable. According to Jeans, Eddington mistakenly believed that "radio-activity is the origin of all the energy radiated from a star" [Jeans, 1917]. In fact, this is not what Eddington stated, so Jeans misinterpreted him. Although Jeans was the first to suggest electron-proton annihilation as a possibility, it was only with Eddington that the idea was applied to the problem of stellar energy [Douglas, 1956, p. 68].[3]

Eddington repeated his suggestion of electron-proton annihilation in a paper of 1919, and the next year he presented a fuller and different version of his ideas of stellar energy in a brilliant address delivered to the annual meeting of the British Association [Eddington, 1919; Eddington, 1920]. Once again he rejected the contraction hypothesis, but now went beyond the annihilation hypothesis by arguing that the energy of a star was released when hydrogen atoms combined to form more complex elements. For the first time Eddington referred to the Rutherford-Bohr model according to which "The nucleus of the helium atom … consists of four hydrogen atoms [protons] bound with two electrons." Might it be possible that helium was formed in the stars by a fusion of electrons and protons?

Eddington thought so and found support in Francis Aston's recent experiments with the new mass spectrograph. Since Aston had found that the mass of a helium nucleus was about 0.7% less than that of four protons, the mass difference $\Delta m$ would be emitted as energy given by $E = \Delta mc^2$. The hypothetical nuclear reaction

$$4\,{}^{1}_{1}\text{H} + 2\,{}^{0}_{-1}e \rightarrow {}^{4}_{1}\text{He}$$

would thus be strongly exothermic. As he wrote: "If 5 per cent of a star's mass consists initially of hydrogen atoms, which are gradually being combined to form more complex elements, the total heat liberated will more than suffice for our demands, and we need look no further for the source of a star's energy." Eddington realised that the hydrogen-to-helium process was conjectural but reminded his audience that Rutherford had recently transmuted nitrogen into carbon by using alpha rays as projectiles.[4] According to Eddington, "what is possible in the Cavendish Laboratory may not be too difficult in the sun."

---

[3] See also [Stanley, 2007a] and [Stanley, 2007b, pp. 58-68] for the dispute between Eddington and Jeans and its basis in their different methodologies. While Jeans favoured a mathematical-deductive method, Eddington's approach was more pragmatic and what he called astronomical.

Eddington realised as the first one that stellar energy generation goes hand in hand with the formation of heavier elements. However, despite his eloquent arguments for the fusion process he did not claim that the true source of stellar energy was to be found in the building up of elements. "It is not," he stated, "of any great consequence whether in this suggestion we have actually laid a finger on the true source of the heat. It is sufficient if the discussion opens our eyes to the wider possibilities." As seen in retrospect, his paper of 1920 is "a milestone in the search for the energy source in stars" [Shahiv, 2009, p. 148], but this is not how it was seen at the time and nor was it how Eddington conceived it.

**3. Eddington's ambivalence**

While Eddington did not mention annihilation in his paper of 1920, in a comprehensive article published in *Zeitschrift für Physik* the following year he discussed both possibilities, namely fusion and annihilation [Eddington, 1921]. For the first time he used Rutherford's new name "proton" for the nucleus of the hydrogen atom. As Eddington pointed out, annihilation was a far more efficient process than fusion, where only about 1% of the mass was transformed into energy. In the first case the total number of electrons and protons would diminish, whereas in the second case the particles would only be rearranged. It took five more years until Eddington returned to the problem of the source of stellar energy, which he discussed in more detail in an article in *Nature* and also in a chapter in his widely read monograph *The Internal Constitution of the Stars*. In both cases he compared "the mild form of transmutation of elements, and the radical form of destruction of matter" [Eddington, 1926b, p. 294; see also Eddington, 1929].

Whatever the form of stellar nuclear energy, there were severe difficulties in relating the rate of energy emission to the temperature and density of the stellar source. Eddington illustrated the difficulty by comparing the Sun's output of energy with the much greater output of the star system Capella. How could this be the case, when it was known that the Sun is hotter and denser than Capella? The Sun was about 600 times as dense, and yet astronomers found the energy output of Capella to be 58 erg/g/s, much more than the 1.9 erg/g/s emitted by the Sun. Instead of addressing the astronomical difficulties, Eddington just evaded them, stating that "there must presumably be some way out of them."

---

[4] Rutherford reported the process to be $^{14}N + ^4He \rightarrow ^{13}C + ^4He + ^1H$, but five years later the products were reinterpreted as $^{17}O + ^1H$. Eddington's version was different and not very clear, namely that Rutherford had "been breaking down the atoms of oxygen and nitrogen, driving out an isotope of helium from them."

Eddington was convinced that nuclear processes were the key to unravel the mystery of stellar energy but undecided of whether fusion or annihilation was the best candidate. As he admitted, both candidates were problematic and in his discussion he focused as much or more on the failures of the two hypotheses as on their successes. After having reviewed the pros and cons of the hypotheses he stated that "we are still groping for a clue" and that "the general result of the arguments is entirely inconclusive" [Eddington 1926b, p. 297]. He was apparently satisfied, at least for the time being, to have glimpsed the obscurity of the problem.

As far as the annihilation hypothesis was concerned Eddington considered it to be "more fertile" to astronomy than the fusion hypothesis, which "leads to no interesting astronomical consequences." He found it unlikely that a proton and free electron should commit "joint suicide" and instead speculated that matter annihilation might take place indirectly by the formation of "certain kinds of nuclei which are self-destroying." He did not elaborate. If Eddington seems to have had a slight preference for the annihilation hypothesis, he also realised that it was purely speculative and unsupported by experiments.[5]

In this respect the fusion alternative fared better, for at least it had a basis in experiments and provided the Sun with a realistic if rather short time-scale. It also suggested an appealing unitary explanation of energy generation and the formation of elements. Moreover, it was "the only process *known* to occur," which is a surprising statement given that there was no experimental evidence at all for the synthesis of helium from hydrogen. In a later version of the stellar energy problem Eddington [1929, p. 101] referred sceptically to "the alleged transmutation of hydrogen into helium in the laboratory," which was most likely a reference to John Tandberg, a Swedish chemist who in 1927 announced to have synthesised helium from hydrogen at room temperature [Kragh, 1996, p. 84].

Among the physical difficulties facing the fusion hypothesis was that it seemed to require an extremely improbable collision of six particles. "How the necessary materials of 4 mutually repelling protons and 2 electrons can be gathered together in one spot, baffles imagination," Eddington wrote. The particles must have enormous velocities and so the interior of the stars must be extremely hot, perhaps hotter than the 40 million degrees allowed by astronomical knowledge. As Eddington noted, from the point of view of nuclear physics, this was "absurdly low [and] … practically at absolute zero." Eddington famously countered the difficulty by means of rhetoric [Eddington, 1926b, p. 301; slightly differently in Eddington, 1926a]:

> It is held that the formation of helium from hydrogen would not be appreciably accelerated at stellar temperatures, and must therefore be ruled out as a source of stellar energy. But the helium which we handle must have been put together at some time and some place. We do not argue with the critic who urges that the stars are not hot enough for this process; we tell him to go and find a *hotter place*.[6]

---

[5] Neither Eddington nor Jeans referred to the "neutron" introduced by Rutherford in 1920 as a tightly bound proton-electron composite. If this particle existed as a stable object, as Rutherford and several other British physicists thought, it would seem to contradict the annihilation hypothesis. When James Chadwick in 1932 discovered the real neutron, at first he interpreted it in Rutherford's sense. Eddington was undoubtedly aware of Rutherford's neutron, but apparently it did not occur to him that it might be of astrophysical relevance.

In 1928 Eddington gave the Fourth Joule Memorial Lecture in Manchester, which he devoted to subatomic energy and its role in astronomy. On this occasion he presented in a popular style what he two years earlier had discussed in his book. Once again he vacillated between the two possibilities of subatomic energy production, both of which were faced with grave difficulties: "We find one fact which seems directly to confirm the hypothesis of annihilation of matter and then another fact goes dead against it. We turn in despair to the other alternative of transmutation of atoms, and that equally leads to trouble" [Eddington, 1928]. However, he cautiously indicated a preference for the annihilation hypothesis: "On the whole, the astronomical evidence is less favourable to this [the fusion alternative] than to the hypothesis of annihilation of electrons and protons, even though the latter involves a greater speculative element." The next year, in his book *Stars and Atoms*, he was less cautious. "Unless we choose annihilation of matter, we cut the life of a star so short that there is no time for any significant evolution at all. … The ultimate particles one by one yield up their energy and pass out of existence. Their sacrifice is the life-force of the stars" [Eddington, 1929, p. 112].

    The problem of stellar energy was well known, not only by astronomers but also by physicists. Rutherford commented on it in an address of 1928 in which he supported Eddington's fusion hypothesis over the idea of pair annihilation. He characterised the latter idea – "which was first put forward by Jeans" – as highly speculative and scarcely worth serious consideration [Rutherford, 1928].

---

[6] The "hotter place" imagined by Eddington was later identified with the big bang during which nearly all the helium in the universe was produced. Of course, Eddington was not thinking of the big bang on something like it, but presumably of some new kind of star. He never accepted the big bang theory of his former student Georges Lemaître.

## 4. Other subatomic hypotheses

The ideas discussed by Eddington and Jeans were not the only suggestions of a subatomic origin of stellar energy. For example, the French physicist and Nobel Prize laureate of 1926, Jean Perrin, discussed in papers of 1920 and 1921 how solar energy might be due to the formation of helium atoms from four hydrogen atoms [Wesemael, 2009]. His ideas were roughly similar to Eddington's but made almost no impact at all among astronomers. Yet Eddington was aware of Perrin's contribution, which he mentioned in some of his works [Eddington, 1926a; Eddington, 1926b, p. 296]. Another Nobel Prize laureate, the German chemist Walther Nernst, disagreed with the fusion hypothesis proposed by Perrin and Eddington. According to him, the radiation produced by the stars was the result of radioactive decay processes rather than the building-up of elements starting with hydrogen.

Based on speculations on a cosmic ether filled with quantum zero-point energy, in the interwar period Nernst developed a theory of a stationary and regenerating universe in which the heaviest elements were the primary material formed by the ether. "The sources of the energy of the fixed stars," he wrote, "must be looked for in radio-active elements which are of higher atomic weight than uranium" [Nernst, 1928; Bartel and Huebener, 2007, pp. 306-25]. It followed from Nernst's view that stellar energy was independent of temperature and pressure. Eddington was from an early date aware of Nernst's unorthodox theory, which he discussed with Nernst but without accepting it [Eddington, 1921]. On the contrary, Eddington flatly dismissed not only Nernst's radioactive theory but also the somewhat similar ideas expounded by Jeans in the same period.

In Jeans' view, proton-electron annihilation was not the only process which contributed to the energy of the stars. He shared with Nernst the predilection for very heavy elements in the stars and the nebulae, and also the view that the stellar bodies evolve from the complex to the simple. In his monograph *Astronomy and Cosmogony* from 1928 and in other publications, Jeans argued that the nebulae and the cores of main sequence stars were rich in transuranic elements with atomic number 95 or more [Kragh, 2013]. The decay of these hypothetical elements supposedly produced the lighter elements and a large part of the energy emitted by the stars. Without referring to either Jeans or Nernst, an American electrical engineer suggested (perhaps with tongue-in-cheek) that stellar energy might be due to the explosive decay of an element with atomic number 118 [Andrews, 1928; Fontani,

9Costa, and Orna, 2015, p. 416]. He called the superheavy element "hypon," possibly a reference to its hypothetical nature.[7]

Jeans' theory was opposed by most physicists and astronomers, who considered it contrived and without empirical justification. After all, no elements with atomic number greater than 92 were known. Eddington was one of the critics. Responding to what he thought were the anti-evolutionary ideas of Nernst and Jeans, he wrote: "Personally, when I contemplate the uranium nucleus consisting of an agglomeration of 238 protons and 146 electrons, I want to know how all these have been gathered together; surely it is an anti-evolutionary theory to postulate that this is the form in which matter first appeared" [Eddington, 1927-28]. Of course, Jeans did not deny cosmic and stellar evolution, he only questioned if evolution necessarily had to go from simple substances to the more complex ones: "It now looks as though the atoms in a star become simple as the star grows older; evolution appears to be from the complex to the simple, and not, as in biology, from simple to complex" [Jeans, 1926].

**5. Eddington's silence**

It seems that at the end of the 1920s Eddington lost interest in the origin of stellar energy. What may have been his last words on the subject appeared in 1935, in a chapter of his popular book *New Pathways in Science*. The major difference from his earlier expositions was that he no longer took the annihilation hypothesis seriously. Eddington considered the recent discovery of the positron to be "a blow to the annihilation hypothesis" and the discovery of the neutron to be yet another blow. Yet he felt it was premature to declare the hypothesis totally dead.

Although Eddington still maintained a cautious attitude, he clearly favoured the fusion hypothesis over the annihilation alternative. Sure, there was still no laboratory evidence for the formation of helium out of hydrogen, "but this objection seems scarcely relevant" [Eddington, 1935, p. 168]. Remarkably, he seems to have been either unaware of or uninterested in the new experiments in Cambridge where Rutherford and his collaborators for the first time reported artificial fusion processes of hydrogen into helium [Oliphant, Harteck, and Rutherford, 1934]. By bombarding heavy hydrogen in the form of ammonium chloride $NH_4Cl$ with deuterons, in 1934 the team at the Cavendish Laboratory interpreted the product to be the hitherto unknown hydrogen isotope H-3 or tritium:

$${}^{2}_{1}H + {}^{2}_{1}H \rightarrow {}^{4}_{2}He^* \rightarrow {}^{1}_{1}H + {}^{3}_{1}H + 4.0 \text{ MeV}$$

---

[7] The existence of element 118 had previously been considered by Niels Bohr and a few other scientists [Kragh, 2013]. Since the manufacture of a few nuclei of this element in 2006, it became real and is today known as oganesson (chemical symbol Og), the heaviest element in the periodic table.

As an alternative, the team suggested that the unstable He-4 nucleus might decay into a He-3 nucleus and a neutron:

$${}^{2}_{1}H + {}^{2}_{1}H \rightarrow {}^{4}_{2}He^* \rightarrow {}^{1}_{0}n + {}^{3}_{2}He + 3.2 \text{ MeV}$$

However, Rutherford did not relate the processes to either stellar energy or element formation in stars. Perhaps for this reason, Eddington did not cite Rutherford and his experiments.

One of Eddington's arguments against annihilation reflected his current work on a new unified theory of microphysics and cosmology [Kragh, 2017a]. As part of this work he reached the conclusion that the number $N$ of protons in the closed universe was a fundamental natural constant, for which he calculated the exact value, namely $N = 136 \times 2^{257}$. A fixed number of protons obviously contradicted the disappearance of a huge number of stellar protons annihilating with electrons. To Eddington – but not to other scientists – this was an additional argument against the annihilation hypothesis.

Although Eddington was preoccupied with his ambitious fundamental theory throughout the 1930s, on a few occasions he also worked on astrophysical problems. Thus, in 1932 he calculated that the hydrogen content in stars was about one third by weight, much more than previously thought [Eddington, 1932]. Hydrogen's predominant role in stars was of importance to the question of energy production, but Eddington ignored the connection. In none of his works after 1930 did he comment on the nuclear-physical approach to stellar energy that eventually led to the celebrated theories of Carl Friedrich von Weizsäcker, Hans Bethe, and Charles Critchfield. It was as if the subject he had pioneered was no longer in his mind. The celebrated breakthrough came with an important conference on "Problems of Stellar Energy-Sources" held in Washington D.C. in 1938 and attended by 34 scientists. I do not know if Eddington was invited, but he did not participate.

The road to our present understanding of the energy sources powering the Sun and other stars is well known and described in the literature [Shahiv, 2009, pp. 275-340; Kragh, 1996, pp. 84-101]. As it turned out, it was quantum mechanics applied to the atomic nucleus that led to progress, an approach which was cultivated by nuclear physicists rather than astronomers. The first fruit was harvested in an important paper by Robert d'Escourt Atkinson and Friedrich Houtermans [1929], and other works in the new nuclear-astrophysical tradition were due to George Gamow, Alan Wilson, and Harold Walke. Progress was far from uniform, though, and for a while stellar energy was considered to be deeply mysterious. Niels Bohr even thought that it indicated a crisis in physics which necessitated radical solutions such as abandoning energy conservation in certain nuclear processes [Kragh, 2017b].

Atkinson and Houtermans suggested in their paper of 1929 that the source of stellar energy was the hydrogen-to-helium fusion process first envisaged by Eddington, and they cited several times Eddington's discussion in *The Internal Constitution of the Stars*. Although in a sense their paper was a vindication of Eddington's old idea, the famous astronomer ignored the Atkinson-Houtermans theory and also later works in the same tradition. He only once referred to Atkinson [Eddington, 1935, p. 179]. So, why did Eddington stay silent when his early ideas of a thermonuclear origin of stellar energy were finally turned into a quantitative theory based on quantum mechanics?

It is hardly a coincidence that Eddington stopped publishing on subatomic stellar energy at about the same time that he initiated his grand cosmo-physical project in which he soon became absorbed. The project greatly appealed to his desire of combining fundamental physics with a new philosophy of nature. These intellectual qualities were absent from the study of stellar energy production and astrophysics generally. Moreover, Eddington was never seriously interested in the new nuclear physics and the pragmatic use of quantum mechanics to understand the nuclear world. In short, he may have lost interest in the stellar energy problem because it was not, to his mind, fundamental enough. All the same, it is surprising that he remained silent and never (as far as I can tell) showed any interest in or commented on Bethe's celebrated theories from the late 1930s.[8]

**Conclusion**

Eddington was undoubtedly one of the pioneers of what became known as nuclear astrophysics, although he and his contemporaries preferred the term "subatomic" over "nuclear." He was the first, or at least one of the first, to advocate particle annihilation and also hydrogen-to-helium synthesis as the processes powering the stars. However, he discussed these hypotheses in a characteristic vague and uncommitted way, and for a long period of time he remained undecided as regards the relative merits of the two hypotheses.

---

[8] [Eddington, 1939] briefly referred to Bethe's theory in relation to white dwarf stars.

Eddington favoured what he called an astronomical approach and regarded the more mathematical approach of the physicists with some suspicion. Without attempting to harvest the fruits of the seeds he had planted, he left the field to the new generation of nuclear physicists, apparently uninterested in their results. It is quite possible that his reluctance and indifference was influenced by his ambitious research project of integrating all of physics in one fundamental and ultimate theory. Eddington's work on this theory was based on a mathematical-deductive methodology completely different from the methods he adopted in his works on stellar energy and astrophysics generally [Stanley, 2007a; Kragh, 2017a]. In any case, although his early work on the sources of stellar energy was brilliant and innovative, it played very little role for the subsequent development. Bethe did not stand on the shoulders of Eddington.